\documentclass{article}

\usepackage{arxiv}

\usepackage[utf8]{inputenc} 
\usepackage[T1]{fontenc}    
\usepackage{hyperref}       
\usepackage{url}            
\usepackage{booktabs}       
\usepackage{amsfonts}       
\usepackage{nicefrac}       
\usepackage{microtype}      
\usepackage{lipsum}
\usepackage{graphicx}
\usepackage{caption}
\usepackage{subcaption}
\usepackage{longtable}
\usepackage{booktabs, multirow} 
\usepackage[toc,page]{appendix}
\usepackage{soul}
\usepackage[table]{xcolor} 
\usepackage{changepage,threeparttable} 
\graphicspath{ {./images/} }

\title{Prediction of drug effectiveness in rheumatoid arthritis patients based on machine learning algorithms}

\author{
 Shengjia Chen \\
  Grossman School of Medicine\\
  New York University\\
  New York, NY 10016\\
  \texttt{sc9295@nyu.edu} \\
   \And
 Nikunj Gupta \\
  Tandon School of Engineering\\
  New York University\\
  New York, NY 11201 \\
  \texttt{ng2531@nyu.edu} \\
  \And
 Woodward B. Galbraith \\
  Center for Data Science\\
  New York University\\
  New York, NY 10011 \\
  \texttt{wbg231@nyu.edu} \\
   \AND
   Valay Shah \\
   Courant Institute of Mathematical Sciences \\
   New York University\\
   New York, NY 10011 \\
   \texttt{valay.shah@nyu.edu} \\
   \And
   Jacopo Cirrone\\
  Center for Data Science\\
  New York University\\
  and Colton Center for Autoimmunity\\
  NYU Grossman School of Medicine\\
  New York, NY 10011 \\
  \texttt{cirrone@courant.nyu.edu} \\
}

\begin{document}
\maketitle
\begin{abstract}
Rheumatoid arthritis (RA) is the most common inflammatory arthritis, affecting 1\% of the population. It is an autoimmune condition resulting in significant joint destruction and morbidity. Machine learning (ML) has the potential to identify patterns in patient electronic health records (EHR) to forecast the best clinical treatment to improve patient outcomes. This study introduced a \textbf{D}rug \textbf{R}esponse \textbf{P}rediction (DRP) framework with two main goals: 1) design a data processing pipeline to extract information from tabular clinical data, and then preprocess it for functional use, 2) predict RA patients’ response to drugs and evaluate classification models' performance. We propose a novel two-stage ML framework based on European Alliance of Associations for Rheumatology (EULAR) criteria cutoffs to model drug effectiveness. In the first stage, the ML models regress the changes in the Disease Activity Score in 28 joints ($\Delta$DAS28) of patients who are bio-naïve to anti-tumor necrosis factor (TNF) treatments; in the second stage, the patient's responses to drugs are classified using predicted $\Delta$DAS28 scores with thresholds. We empirically show that such division into subtasks significantly improves the accuracy of predicting drug effectiveness in RA patients. Furthermore, regression of $\Delta$DAS28 scores makes our model more interpretable to health care providers, and the classification of the change between initial and 3-month DAS scores would give an easy-to-understand binary recommendation. Our model Stacked-Ensemble DRP was developed and cross-validated using data from 425 RA patients. The evaluation used a subset of 124 patients (30\%) from the same data source. In the evaluation of the test set, two-stage DRP leads to improved classification accuracy over other end-to-end classification models for binary classification. Our proposed method provides a complete pipeline to predict disease activity scores and identify the group that does not respond well to anti-TNF treatments, thus showing promise in supporting clinical decisions based on EHR information. The code is open source and is available on GitHub: \url{ https://github.com/Gaskell-1206/Ensemble_DRP}.
\end{abstract}


\section{INTRODUCTION}
\label{sec:Introduction}
Rheumatoid arthritis (RA) is an autoimmune disease that primarily attacks the synovial tissues within the joints, commonly affecting joints in the hands, knees, and wrists. RA causes inflation in the lining of effected joints, potentially leading to chronic pain, unsteadiness, and deformity. The annual incidence of RA in the United States is approximately 40 per 100,000 persons \cite{chaudhari2008impact}. The lifetime likelihood of developing RA for U.S. adults is 3.6\% for women and 1.7\% for men \cite{crowson2011lifetime}.

The diagnosis of RA involves symptom review, physical examination, X-rays, and lab tests. The best treatment result for RA is if diagnosis occurs within six months of the patient first developing symptoms. The American College of Rheumatology (ACR) recommends assessment tools to determine the disease stage of RA patients. The standard treatments for RA are two classes of disease-modifying antirheumatic drugs (DMARDs): 1) traditional DMARDs such as methotrexate, and 2) biologic DMARDs such as tumor necrosis factor (TNF) inhibitors and non-TNF DMARDs. Comparisons of the two classes indicate that biologics are more potent than traditional DMARDs for most patients \cite{nurmohamed2008biologics}. However, serious side effects of biologic DMARDs include increased risk of opportunistic infections and reactivation of latent tuberculosis. Furthermore, biologic RA treatment drugs are expensive, costing up to \$20,000 annually per patient \cite{chaudhari2008impact}. The biologic DMARDs have a wide variation in efficacy. For example, TNF inhibitors have little or no effect on approximately 30\% of patients, which leads to unnecessary costly drug use and unimproved disease conditions.

Therefore, there are clinical and financial advantages to developing a consistent and accurate predictive model that can provide a prospective prediction of the response of individual patients to various biologic DMARDs. Predicting the response of RA patients to specific anti-TNF treatments poses a challenge for advancing precision medicine to this specialty. Accurate predictions can provide valuable information on drug selection, thus helping patients to avoid serious side effects, unnecessary monetary expenditure, and time delays beyond the six months when medication is the most effective. However, conventional models have failed to provide accurate predictions with limited features due to the heterogeneity of RA patient responses. In this study, we aim to develop, validate, and compare state-of-the-art ML models for the prediction of biological drug effectiveness in RA patients.

Using data about patient demographics, baseline disease assessment, lifestyle factors, medical history, lab test results, and treatment from the Consortium of Rheumatology Researchers of North America (CORRONA) - Comparative Effectiveness Registry to Study Therapies for Arthritis and Inflammatory Conditions (CERTAIN), we created a stacked ensemble regression model for a two-stage approach: Stage 1 to predict changes in the Disease Activity Score in 28 joints (DAS28), and Stage 2 to classify patients into either the responder or the non-responder group. This model was developed and cross-validated by a training set of 425 RA patients. The final evaluation used a test data set of 124 patients from the CORRONA CERTAIN. We also compared this two-stage approach with direct classification models.

The main contributions of this paper are the use and evaluation of several machine learning methods on electronic health record data for predicting the effectiveness of drugs in RA patients. We also proposes novel two-stage learning approach to achieve better performing models. Section \ref{sec:Previous Work} discusses previous works predicting the response of patient with RA followed by a description of CERTAIN dataset used in Section \ref{sec:Datasets}. Then, in Section \ref{sec:Methodology}, we discuss in detail the data preprocessing (Section \ref{sec:Data Preprocessing}), the ML baselines (Section \ref{subsec:model family}), and the novel two-stage learning approach to obtain better predictions (Section \ref{subsec:two stage model}). Lastly, Section \ref{sec:Experiments and Results} provides a detailed study of ML methods (comparing baseline and three-month predicted results) used in this study and evaluation metrics that compared all the learned models.

\section{PREVIOUS WORK}
\label{sec:Previous Work}
Initial work predicting the response of patients with RA to tumor necrosis factor (TNF) inhibition mostly applied linear regression modeling to identify the primary factors affecting the likelihood of a positive patient response \cite{anderson2000factors}. Iwamoto et al. (2009). \cite{iwamoto2009prediction} used logistic regression models to find the baseline variables in patients with RA treated with etanercept, an anti-TNF biologic that can induce remission. Several studies have found multiple predictors of response to anti-TNF drugs, such as high serum CRP level \cite{westhovens2014early}, rheumatoid factor (RF) \cite{klaasen2011value}, sex, age \cite{kleinert2012impact}, and BMI \cite{gremese2013obesity, klaasen2011body}. However, the computational task of predicting drug response remains challenging due to limitations on data availability and algorithmic shortcomings \cite{adam2020machine}. Recently, researchers have focused on applying powerful and robust ML algorithms to develop a personalized approach for RA treatment. Convergent Random Forest (CRF) is used to find highly predictive biomarkers capable of predicting anti-TNF response using gene expression data \cite{bienkowska2009convergent}. Miyoshi et al. (2016) \cite{miyoshi2016novel} used a neural network with only nine clinical variables to predict the clinical response to infliximab (IFX). Guan et al. (2019) \cite{guan2019machine} proposed a Gaussian process regression (GPR) model that integrated both clinical and omics biomarkers from the Dialogue on Reverse Engineering Assessment and Methods (DREAM): Rheumatoid Arthritis Responder Challenge \cite{plenge2013crowdsourcing, sieberts2016crowdsourced} to predict the response of patients with RA to TNF inhibitors.

In our work, we adopted a two-stage model that incorporated six base ML model families to predict rheumatoid arthritis patients’ response to anti-TNF treatments. Two-stage methods for machine learning study have been studied in various fields. Hwang et al. (2020) \cite{hwang2020two} built a two-stage mixed ransomware detection model combined with the Markov model and Random Forest. Constantine et al. (2003) \cite{constantine2003improved} proposed a sensitive/less sensitive test strategy including infection and then an enzyme immunoassay to classify HIV-1 infection. The two-stage machine learning model considers domain knowledge, contributing to model interpretability and better classification accuracy. In the current study, we used EULAR criteria as classification cutoffs to categorize changes of DAS28 to facilitate rapid identification of RA patients' response to drugs and coupled its use with a stacked ensemble machine learning model to improve model performance.

\section{DATASETS AND SUMMARY STATISTICS}
\label{sec:Datasets}
The study was conducted using electronic health records (EHR) data gathered from CERTAIN. CERTAIN is a substudy of CORRONA, which has collected clinical RA Data since 2002. CORRONA (now renamed CorEvitas) aims to provide longitudinal, long-term, real-world data for the rheumatology research community \cite{sieberts2016crowdsourced}. CERTAIN recruited patients from the existing CORRONA network; those who fulfill the 1987 ACR criteria of moderate disease or more advanced disease activity \cite{arnett1988american}. The data set contains 2,814 patients across 43 sites. Prior to the start of drug treatment, each patient completes a baseline evaluation that collects an array of biosamples (including DNA, RNA, plasma, and serum). Following this evaluation, each patient’s laboratory blood data, visits, outcome measures, and physician's evaluation are collected every three months (quarterly) for a year \cite{pappas2014design}.

All study data were extracted from CORRONA CERTAIN. We considered all adult patients (age $\geq$ 18) with RA who 1) fulfilled the 1987 ACR criteria, 2) had at least moderate disease activity defined by clinical disease activity index (CDAI) score > 10, and 3) who began or switched biologic agents \cite{aletaha2005remission}. Furthermore, all patients had a baseline (the first onset visit or month 0) DAS28 > 10. All selected patients were anti-TNF-naïve (i.e., no record of using TNF inhibitors prior to the baseline study period) because the study of biological-naïve patients can provide more useful information for clinicians and consistent data distribution for machine learning algorithms while more features are required to study treatment response for biological-experienced patients. Applying these criteria, we identified 425 patients out of a total of 1,229 patients in the CORRONA CERTAIN database. In our study, only baseline features were used as predictors of DAS28 in the third month since rheumatologists are more intended to know patients’ treatment responses within the first three months and most anti-TNF have a relatively rapid onset of action at therapeutic doses in 6-8 weeks.

\begin{table}[!ht]\centering
\label{Summary statistics}
\caption{Summary Statistics for key predictors, treatments, and prediction targets in training and test set}
\begin{tabular}{c|c|c|c}
\rowcolor[HTML]{A6A6A6} 
Groups                            & Features                           & Training   (n=257) & Test   (n=120) \\
                                  & Age (years), mean (SD)             & 54.75 (12.78)      & 55.01 (13.07)  \\
\multirow{-2}{*}{Demographics}    & Sex (female), count                & 189 (73.58\%)      & 91 (75.81\%)   \\
\midrule
                                  & Humira/adalimumab,   count         & 98 (38.13\%)       & 51 (42.50\%)   \\
                                  & Enbrel/Etanercept, count           & 79 (30.74\%)       & 34 (28.33\%)   \\
                                  & Remicade/infliximab,   count       & 55 (21.40\%)       & 23 (19.17\%)   \\
                                  & Cimzia/certolizumab pegol,   count & 23 (8.95\%)        & 6 (5.00\%)     \\

\multirow{-5}{*}{Treatment}       & Simponi Aria/golimumab, count      & 2 (0.778\%)        & 6 (5.00\%)     \\
\rowcolor[HTML]{A6A6A6} 
\multicolumn{4}{c}{\cellcolor[HTML]{A6A6A6}Targets}                                                          \\
                                  & Baseline DAS28, mean (SD)          & 4.98 (1.03)        & 4.82 (0.90)    \\
\multirow{-2}{*}{DAS28 (3M)}      & $\Delta$DAS28, mean (SD)                  & 1.55 (1.24)        & 1.40 (1.40)    \\
                                  & Responder, count                   & 196 (76.26\%)      & 87 (72.5\%)    \\
\multirow{-2}{*}{Response Status} & Non-Responder, count               & 61 (23.74\%)       & 33 (27.5\%) \\
\bottomrule
\end{tabular}
\begin{tablenotes}
{\scriptsize
  \item[*] Note: This table summarizes statistics of the training and test sets for this study. The test set is created by selecting samples without missing values in target features used for calculating changes of DAS28 at three months.}
\end{tablenotes}
\end{table}

\section{METHODOLOGY}
\label{sec:Methodology}
\subsection{Data Preprocessing}
\label{sec:Data Preprocessing}
\paragraph{Features/Predictors for training.} We extracted clinical data from electronic health records as predictors (i.e., features or variables used in the machine learning models). Such clinical information can be categorized into seven groups of predictors and are summarized as follows:
\begin{itemize}
    \item{Patient demographics}: age, gender, race, ethnicity, body mass index (BMI)
    
    \item{Lifestyle factors}: smoking history and status, alcohol consumption
    
    \item{Patient groups}: drug group, initial patient group (bionaïve or bioexperience)
    
    \item{Medication exposures history}: \#TNF, \#nonTNF, NSAIDs, methotrexate cotherapy(MTX)
    
    \item{Comorbidity}: diabetes, cancer, stroke, COPD, CHF, TIA
    
    \item{Patient Assessment}: tender joints counts(TJC), swollen joints counts(SJC), patient global assessment(GH), patient pain(pt\_pain) 
    
    \item{Lab test}: C-reactive protein (CRP) level, IgA, IgG, IgM
\end{itemize}

\paragraph{Feature Engineering.} To convert raw data into a more usable format, features were dropped in the following ways. First, features that provided redundant information were dropped, such as visit date, visit time, and time between visit date and onsite date. Dropping features was useful for having a low number of features, consequently reducing model complexity. Second, features with more than 70\% missing values were dropped. Imputing so many missing values holds a higher chance of inducing bias. Third, features that were constant (i.e., had a standard deviation of zero) were dropped. Additionally, a label encoder was implemented to convert categorical features into numeric features for ease of training our ML models.

\paragraph{Missing Value Imputation.} A summary of missing data for each attribute and performance comparison of each imputation method can be found in Appendix \ref{performance imputation} and \ref{feature dictionary}. Due to the low quantity of the data used in this study, it was not feasible to drop individuals with minor missing values, hence making the imputation of such values necessary. Now, improper imputation of these values could bias our study; we carefully considered several imputation approaches; k-nearest neighbor (KNN) imputation was the ultimate choice.

Features that do not change over time were imputed based on the last observed value for an individual. As also mentioned, to further avoid bias, only individuals with complete data for features relevant to DAS-28 at three months were used in the test set.

\paragraph{Handling Data Imbalance.} 
During the model training, we found that most data belong to responders who had 223 rows in the training set, around 77\% of the total samples. The unbalance class might lead to overfitting problems when an ML model attempts to predict the major class with higher accuracy but fails in predicting the minority class. Therefore, we considered a combined scenario to overcome the unbalance problem: 1) For the training set, we undersampled the majority class and oversampled the minority class, 2) For cross-validation, we adopted repeated stratified K Fold to generate validation sets that contain the same distribution of classes, and 3) When training models, cost-sensitive learning and ensemble learning models were used to learn equally from each class by using a class weight.

\paragraph{Feature to Regress: $\Delta$DAS28 Score.} 
To integrate different aspects of disease activity in patients with RA and comprehensively assess patients’ response to treatment, DAS is used as a combined index \cite{van1990judging}. DAS has been extensively validated for its use in clinical trials in combination with the EULAR response criteria \cite{fransen2004disease}. DAS28, a derivative of DAS and more commonly used, assesses 28 joints, with a score ranging from 0 to 10 calculated by the number of swollen and tender joints, patient’s global assessment, and a laboratory measure of acute inflammation which can be the C-reactive protein (CRP) or erythrocyte sedimentation rate (ESR) \cite{fransen2004disease}. The formula for calculating DAS28 is as follows: 
\begin{equation}
    \Delta DAS28 = ( 0.56 * \sqrt{TJC} ) + (0.28 * \sqrt{SJC}) + ( 0.36 * \ln{(CRP + 1)} ) + (0.014 * GH) +0.96
\end{equation}
where TJC is the number of tender joints of 28 counted, SJC is the number of swollen joints of 28 counted, CRP is the C-reactive protein level in mg/liter, and GH is the patient's global assessment on a 100-mm VAS.

Based on its properties, stakeholders, health systems, and payers often use DAS28 in determining the need for therapeutic regimens and clinicians can use it to correctly choose treatment decisions the first time, saving precious time that would otherwise be wasted trying and failing with other treatments. In addition, clinicians can use it to evaluate patients’ treatment outcomes \cite{greenmyer2020das28}.

\paragraph{Classes to predict.} Categorization is important for clinical decision support about entry into clinical trials as well as requirements for therapeutic changes and for setting therapeutic goals. The EULAR criteria have been proposed and used for the original DAS and the DAS28 to identify high and low levels of RA activity. In this study, we used the EULAR response criteria defined in Table \ref{table:classify labels} to classify DAS28 scores. By calculating the changes of DAS28 between baseline and three months, it is possible to define improvement or patients’ response to anti-TNF drugs. In the first stage experiment, we considered a binary classification task: 1) patients with ’Good’ and 'Moderate' responses would be grouped into positive samples as 'Responder', and 2) patients with 'No Response' would be turned into 'Non-Responder'. To solve the problem of imbalance dataset, we adopted a three-class classification task in the second stage experiment. In three classes, the majority class is divided into two subclasses, which not only contributes to model training but also provide more detailed information for doctors.

\begin{table}[htbp]
\centering
\caption{Classification cutoff of Drug Response according to EULAR criteria for end and changes of DAS28}
\resizebox{0.8\columnwidth}{!}{%
\begin{tabular}{l*{3}c}
\toprule
    & \multicolumn{3}{c}{DAS28} \\
    \cmidrule(r){2-4}  
    $\Delta$DAS28 & \textbf{D} $\leq$ 3.2 & $3.2<$ \textbf{D} $\leq 5.1$ & \textbf{D} $>$ 5.1 \\
\midrule
$\Delta \leq 0.6$ & No Response & No Response & No Response \\
$0.6<\Delta \leq 1.2$   & Moderate & Moderate & No Response \\
$\Delta > 1.2$  & Good  & Moderate & Moderate               \\
\bottomrule
\end{tabular}
}
\begin{tablenotes}
{\scriptsize
  \item[*] Note: $\Delta$ represents $\Delta$DAS28 which is defined by $\Delta$DAS28 = (baseline DAS28 – three-months DAS28), D represents DAS28 of the end month (three months). Different colors denote three different classes.}
\end{tablenotes}
\label{table:classify labels}
\end{table}

\subsection{Machine Learning Model Family (Baselines)}
\label{subsec:model family}

Initially, we implemented several baseline machine learning models for this study. The model families evaluated were generalized linear models (GLMs) \cite{mccullagh2019generalized}, multi-layer artificial neural networks (ANNs) \cite{lecun2015deep}, Distributed Random Forest (DRF) \cite{ho1995random}, Extremely Randomized Trees (XRT) \cite{geurts2006extremely}, Gradient Boosting Models (GBM) \cite{natekin2013gradient}, and Extreme Gradient Boosting (XGB) \cite{chen2016xgboost}.

Now, since the ensemble machine learning methods have been extremely successful in many domains, we investigated if their adoption further improves the performance of prediction and classification of drug effectiveness in RA patients. Ensemble learning is a technique of combining the predictions of multiple classifiers (or regressors) to produce a single classifier (or regressor). The resulting classifier (or regressor) is generally more accurate than any of the individual classifiers (or regressors) comprising the ensemble. In this study, we used three approaches to ensemble training: bagging \cite{breiman1996bagging}, boosting \cite{freund1996experiments,schapire1990strength}, and stacking \cite{van2007super}. Bagging considers homogeneous weak learners, learns them independently from each other in parallel and combines them following a deterministic averaging process. Boosting learns them sequentially in a very adaptative way (a base model depends on the previous ones) and combines them following a deterministic strategy. Whereas, stacking, considers heterogeneous weak learners, learns them in parallel and combines them by training a second-level meta-model to output a prediction based on the different weak models predictions. \footnote{We used h20 \cite{ledell2020h2o} and Scikit-learn \cite{pedregosa2011scikit} python packages to implement all the mentioned algorithms.} 

As expected, stacking ensemble models outperformed the other baselines including bagging ensemble and boosting ensemble (discussed further in detail in Section \ref{sec:Experiments and Results}). Nevertheless, for our use case, we found that sub-dividing the classification of the drug effectiveness in RA patients into two stages led to improved performance. In the next section (Section \ref{subsec:two stage model}), we describe in detail, our novel approach to achieve better model performance to classify drug effectiveness in RA patients.

\subsection{Two-stage Model}
\label{subsec:two stage model}
\begin{figure}[!ht]
  \centering
  \includegraphics[width=0.8\linewidth]{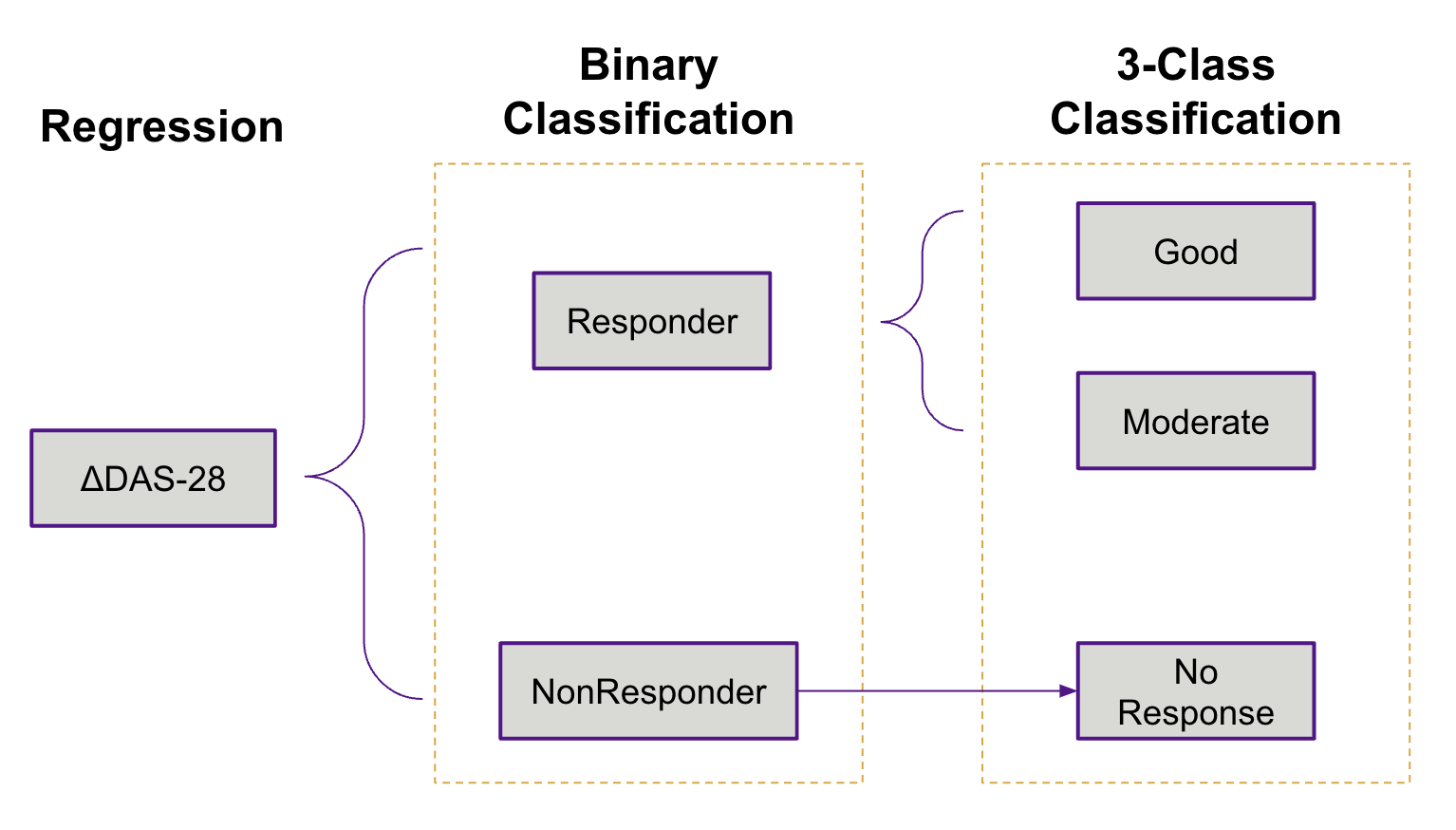}
  \caption{Two-stage approach for the treatment response prediction model. DAS-28 CRP is calculated by four target features, and predicted by regression models and then used to identify nonresponders using thresholds suggested by EULAR criteria.}
  \label{fig:two stage approach pipeline}
\end{figure}

We propose a two-stage approach for learning to classify the drug effectiveness in RA patients: 1)build a regression model to predict the changes of DAS28 score ($\Delta$DAS28), and 2) use this predicted $\Delta$DAS28 as input to classify the treatment response. A complete pipeline can be found in Figure \ref{fig:two stage approach pipeline}.

In both clinical trials, the DAS28 can be used to assess whether an individual patient has a significant improvement of the disease activity, compared to baseline. In addition, the DAS28 can also be helpful in clinical practice. Treatment decisions can be based on current DAS28 values or on changes in DAS28 compared to values before the start of the treatment. This approach has several advantages. Firstly, the DAS28 and its thresholds for high and low disease activity have been extensively validated \cite{fleischmann2017das28,greenmyer2020das28, matsui2007disease}. Moreover, a clear relationship exists between the mean DAS during a certain period and the amount of radiographic damage developed by the patient in that timeframe \footnote{\url{https://www.das-score.nl/en/}} 3. Furthermore, we also show in this article that on applying ML algorithms, the validated threshold used in two-stage approach outperforms optimized classification threshold. The classification results can be more straightforward to compare the model performance using prediction accuracy.

\section{Experiments and Results}
\label{sec:Experiments and Results}

In this section, we will present, describe, and discuss the experiments and corresponding results for the ML baseline models mentioned earlier \ref{subsec:model family} and our novel two-stage approach learning model \ref{subsec:two stage model}.

\subsection{Comparison of all the models in both stages}
\label{sec:Comparison of all the models in both stages}

\begin{figure}[!ht]
  \centering
  \includegraphics[width=0.8\linewidth]{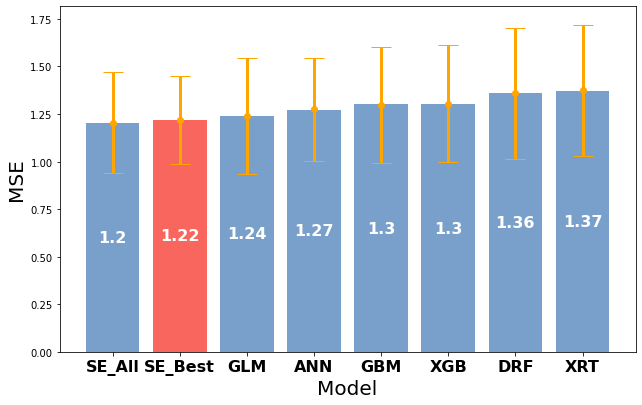}
  \caption{Error bar graph comparing Mean Squared Errors (MSE) between the ground truth of change in DAS28 and predictions from tested regression models. The abbreviations for the terminology in the error bar indicated in this figure stand for the following: Stacked Ensemble model with all models (SE\_All), Stacked Ensemble model with best models in each base model family (SE\_Best),  Generalized Linear Models (GLM), Artificial Neural Networks (ANNs), Gradient Boosting Models (GBM), Extreme Gradient Boosting (XGB), Distributed Random Forest (DRF), Extremely Randomized Tree (XRT).}
  \label{fig:regression CV mse}
\end{figure}

\begin{figure}[!ht]
     \centering
     \begin{subfigure}[b]{0.48\textwidth}
         \centering
         \includegraphics[width=\textwidth]{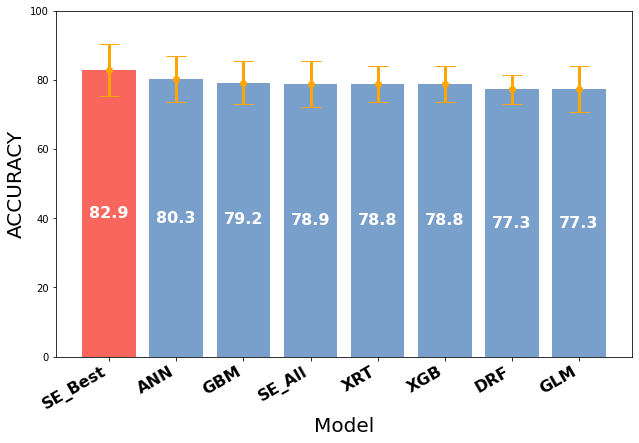}
         \label{fig:classification acc}
     \end{subfigure}
     \hfill
     \begin{subfigure}[b]{0.48\textwidth}
         \centering
         \includegraphics[width=\textwidth]{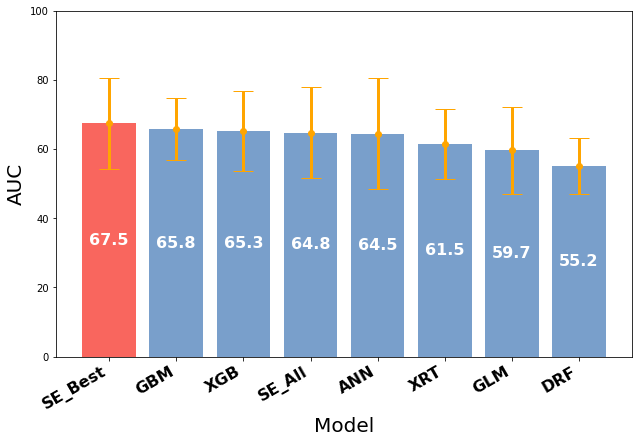}
         \label{fig:classification auc}
     \end{subfigure}
     \caption{(a) Error bar for classification accuracy (\%), between the ground truth of drug response and predictions from tested classification models. (b) Error bar for classification area under ROC (AUC), between the ground truth of drug response and predictions from tested classification models. AUC is scaled to [0,100].}
     \label{fig:classification performance}
\end{figure}

By using a total of 55 features of tabular medical information as described in Table \ref{feature dictionary} (See the Appendix) and the two-stage approach, we predicted $\Delta$DAS28 for each patient (first stage) and classified them as \textit{``responders''} or \textit{``non-responders''} as shown in Figures \ref{fig:regression CV mse} and \ref{fig:classification performance}. The prediction models were evaluated using 3-round 10-fold stratified cross-validation. Each round of the repeated tests started by randomly dividing the training data into 10-folds of equal size. Each fold had the same ratio of instances of target variable as in the whole training set, which  resolved the imbalanced data problem.

\paragraph{First Stage: Regression of $\Delta$DAS28 score.} We evaluated the stacked ensemble models through cross-validation tests and compared them with the base model families (GLM, GBM, DRF, XRT, deeplearning, xbgboost). For $\Delta$DAS28 score prediction (regression task), stacked ensemble with all models achieved the best average mean squared error (MSE) of 1.20, followed by the stacked ensemble model with best model of each model family (1.22) and GLM (1.19) (Figure \ref{fig:regression CV mse}). In model selection, a total of 30 scores generated by 10 rounds of cross-validation was averaged for the estimated model performance score. The $\Delta$DAS28 for regression task was predicted by regression models and evaluated using mean squared error (MSE). Due to the small number of data points, we also took the standard deviation into consideration. Therefore, the model with the lowest variance among top 3 models is selected as the final model: the stacked ensemble model with best model of each model family. The generalized linear model (GLM) was used both as single base model and the meta learner in the second level of stacked ensemble models. The single models in the GLM family shared properties with stacked ensemble models and had relatively similar performance in regression tasks. Applying stacked ensemble to the same predictors increased MSE ranging from 0.04 to 0.15, respectively.

\paragraph{Second Stage: Classification of predicted $\Delta$DAS28 score.} In terms of drug response binary classification, stacked ensemble model with the best of each model family reached the best performance with average weighted F1-score of 88.8 and AUC score of 0.675. Thus, it was chosen as the final model in the test evaluation. The model families with core idea of boosting, such as gradient boosting models (GBM) and Extreme Gradient Boosting (XGB), turned out to be better classifiers compared to other base model families in our experiments (Figure \ref{fig:classification performance}). Compared to base line random classification (i.e., 50\% chance of binary classification accuracy), using a stacked ensemble models increased the rate of accurate classification by 32.9\%.

\subsection{Evaluation on test set}
\label{subsec:Evaluation on test set}

We evaluated the best model selected in cross-validation on the test set, which consisted of information on 124 patients (30\%) without any missing target variable values. The stacked ensemble model trained on the regression task was evaluated using a two-stage approach (i.e., classify patients into two groups using EULAR criteria), achieved an MSE of 1.19 when predicting $\Delta$DAS28 score and classification accuracy of 75.0\% (and weighted F1 score of 0.701) when classifying anti-TNF non-responders. On the other hand, the stacked ensemble model, which was trained directly for the classification task, could achieve an accuracy of only 67.5\% (and the weighted F1 score of 0.64). This gap in the accuracy of 7.5\% between the two methods in test evaluation indicating the proposed two-stage approach with the stacked ensemble model not only provided predictable $\Delta$DAS28 values but also accurately classified patients’ responses to anti-TNF. However, the model performance of XGB and XRT also reach the same classification accuracy but a lower F1 score compared to the stacked ensemble and better F1 score in directly classifying patients, indicating that a powerful single base model could overfit the majority class and thus return an opposite result. 

\begin{table*}[htbp]
\centering
\caption{Model performance compared two-stage approach and direct classification in evaluation on the test set.}
\begin{tabular}{c|c|cc|cc}
\toprule
\multicolumn{2}{c}{Metrics} & \multicolumn{2}{c}{Accuracy} & \multicolumn{2}{c}{F1} \\
\midrule
Model Class & Model & Two-stage & Classification & Two-stage & Classification \\
\midrule
& SE\_Best & \textbf{0.750}                                 & 0.675          & {\textbf{0.701}} & 0.640          \\
\multirow{-2}{*}{\begin{tabular}[c]{@{}c@{}}Stacked\\ Ensemble\end{tabular}}  & SE\_ALL  & 0.725                                 & 0.733          & 0.676                                 & 0.642          \\
\midrule
                                                                              & GBM      & 0.725                                 & 0.658          & 0.609                                 & 0.606          \\
\multirow{-2}{*}{\begin{tabular}[c]{@{}c@{}}Boosting\\ Ensemble\end{tabular}} & XGB  & {\textbf{0.750}} & 0.733          & 0.691                                 & 0.702          \\
\midrule
                                                                              & XRT      & \textbf{0.750}                                 & 0.742          & 0.664                                 & 0.647          \\

\multirow{-2}{*}{\begin{tabular}[c]{@{}c@{}}Bagging\\ Ensemble\end{tabular}}  & DRF      & 0.733                                 & 0.717          & 0.654                                 & 0.619          \\
\midrule
Deep Learning                                                                 & ANN      & 0.742                                 & 0.508          & 0.695                                 & 0.530          \\
Linear Model                                                                  & GLM      & 0.717                                 & 0.717          & 0.619                                 & 0.619         \\
\bottomrule
\end{tabular}
\centering
\begin{tablenotes}
{\scriptsize
  \item[*] Note: The best model is selected in each model family (including stacked ensemble) by the lowest standard deviation among the top-3 mean accuracy. For two selected evaluation metrics, both the two-stage approach and the direct classification model are compared for the binary classification task. The best performance in each column is highlighted indicating that the two-stage approach improves overall classification performance.}
\end{tablenotes}

\end{table*}

\subsection{Ablation Study: Treatment response prediction on 3 classes }
\label{subsec:Treatment response prediction on subclass}

Due to limited data, we could only validate and test models on data with imbalanced classes. To solve this problem and explore a subclass of the responder class, we divided responders into “good” and “moderate” based on the updated EULAR criteria. The same evaluation metrics and model selection criteria were used to evaluate the models' performance. In this ablation study, the two-stage approach outperformed 3-class direct classification models. The stacked ensemble model with the best model in each family reaches 55.8\% accuracy (compared to 33.3\% as the baseline – random classifier), followed by ANNs (53.3\%) and the stacked ensemble model with all models (49.2\%). Nonetheless, the base models outperformed stacked ensemble models in the direct classification tasks. In a more complicated multiclass classification task, the stacked ensemble models overfit to major samples in the training set but fail to correctly predict minority class using a classification model. The best model turned out to be XGB which could reach 49.3\% accuracy (compared to 33.3\% as the baseline -- random classifier), followed by gradient boosting model family (46.3\%) and ANNs (43.5\%) as the next best performing models.

\section{Conclusion}
\label{sec:Conclusion}
Our two-stage approach with the stacked ensemble method led to substantially improved discrimination and calibration for predicting the RA patients’ response to anti-TNF treatments. Model performance remained stable across a range of cross-validation and external validation. Although the stacked ensemble model does not outperform all single base models in classification accuracy, it can still reach the best F1 score which is most important in an unbalance classification task. Most importantly, the two-stage approach improves overall model performance by incorporating verified domain knowledge. These findings support the potential advantages of incorporating ML models into clinical decision-making by providing drug selection suggestions.

Our framework can be easily adapted to other disease cohorts or similar drug response prediction tasks. Moreover, several directions for future work have been identified: 1. Interpret stacked ensemble models by aggregating feature importance from each single base model, 2. Expand the dataset to investigate bio-experienced patients, and 3. Since we used the EULAR criterion to classify patients' responses in the second stage of our proposed ML framework, a promising future study could use ML for the second stage and further investigate 3-Class classification.

\bibliographystyle{unsrt}  
\bibliography{references}  

\newpage
\begin{appendices}
\section{Performance Comparison for Imputation Methods}
\label{performance imputation}
\begin{figure}[!ht]
  \centering
  \includegraphics[width=0.75\linewidth]{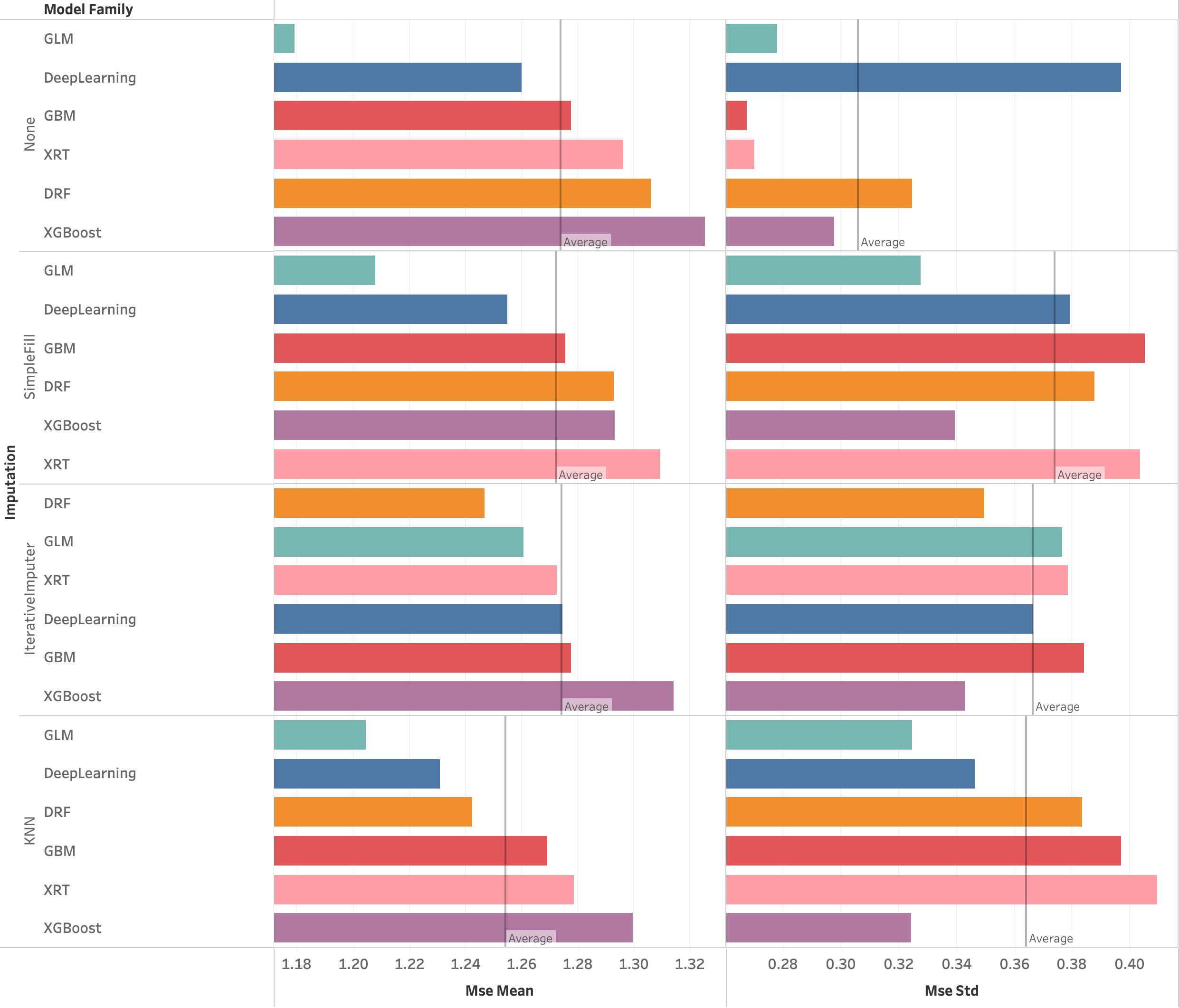}
  \caption{Performance comparison for imputation methods. ‘None’ denotes no imputation is applied. Three imputation methods are compared using six base model families. The best imputation method is selected by the lowest average MSE.}
  \label{fig:comparison imputation}
\end{figure}

\section{Feature Dictionary}
\label{feature dictionary}
\begin{longtable}{|l|l|l|l|l|l|}
\caption{Feature dictionary and missing value summary for training and test set. Feature name is consistent with the CORRONA CERTAIN dataset. NaN value in mean represents that the feature is categorical.}\label{tab:Feature Dictionary} \\
\hline \multicolumn{1}{|c|}{\textbf{Feature Group}} & \multicolumn{1}{c|}{\textbf{Feature}} & \multicolumn{1}{c|}{\textbf{Train Mean}} & \multicolumn{1}{c|}{\textbf{Train Null}} &
\multicolumn{1}{c|}{\textbf{Test Mean}} &
\multicolumn{1}{|c|}{\textbf{Test Null}} \\ \hline 
\endfirsthead

\multicolumn{5}{c}%
{{\bfseries \tablename\ \thetable{} -- continued from previous page}} \\
\hline \multicolumn{1}{|c|}{\textbf{Feature Group}} & \multicolumn{1}{c|}{\textbf{Feature}} & \multicolumn{1}{c|}{\textbf{Train Mean}} & \multicolumn{1}{c|}{\textbf{Train Null}} &
\multicolumn{1}{c|}{\textbf{Test Mean}} & \multicolumn{1}{|c|}{\textbf{Test Null}} \\ \hline 
\endhead

\hline \multicolumn{5}{|r|}{{Continued on next page}} \\ \hline
\endfoot

\hline \hline
\endlastfoot

\multirow{2}{*}{\textbf{Patient Group}} &grp &NaN &0 &NaN &0 \\\cmidrule{2-6}
&init\_group &NaN &0 &NaN &0 \\\cmidrule{1-6}
\multirow{8}{*}{\textbf{Demographics}} &age &54.75 &0 &55.01 &0 \\\cmidrule{2-6}
&gender &NaN &0 &NaN &0 \\\cmidrule{2-6}
&final\_education &NaN &1 &NaN &0 \\\cmidrule{2-6}
&race\_grp &NaN &0 &NaN &0 \\\cmidrule{2-6}
&ethnicity &0.92 &6 &0.94 &2 \\\cmidrule{2-6}
&weight &186.55 &1 &180.17 &0 \\\cmidrule{2-6}
&BMI &30.12 &1 &29.19 &0 \\\cmidrule{2-6}
&height &65.86 &0 &65.85 &0 \\\cmidrule{1-6}
\multirow{3}{*}{\textbf{Lifestyle}} &newsmoker &NaN &8 &NaN &3 \\\cmidrule{2-6}
&drinker &NaN &0 &NaN &2 \\\cmidrule{2-6}
&drinksperwk &3.07 &0 &3.17 &2 \\\cmidrule{1-6}
\multirow{5}{*}{\textbf{RA History}} &duration\_ra &4.84 &0 &5.38 &0 \\\cmidrule{2-6}
&ara\_func\_class &NaN &1 &NaN &0 \\\cmidrule{2-6}
&num\_tnf &0 &0 &0 &0 \\\cmidrule{2-6}
&num\_nontnf &0 &0 &0 &0 \\\cmidrule{2-6}
&ndmardused &1.5 &0 &1.65 &0 \\\cmidrule{1-6}
\multirow{4}{*}{\textbf{Clinical Values}} &rfstatus\_b &0.64 &0 &0.68 &0 \\\cmidrule{2-6}
&usresultsRF &121.37 &1 &132.76 &0 \\\cmidrule{2-6}
&ccpstatus\_b &0.48 &0 &0.6 &0 \\\cmidrule{2-6}
&usresultsCCP3 &125.31 &9 &135.23 &3 \\\cmidrule{1-6}
\multirow{11}{*}{\textbf{Comorbidity}} &hxmi &0.01 &0 &0.02 &0 \\\cmidrule{2-6}
&hxunstab\_ang &0 &0 &0 &0 \\\cmidrule{2-6}
&hxchf &0 &0 &0.02 &0 \\\cmidrule{2-6}
&hxother\_cv &0.03 &0 &0.01 &0 \\\cmidrule{2-6}
&hxtia &0 &0 &0 &0 \\\cmidrule{2-6}
&hxstroke &0.01 &0 &0 &0 \\\cmidrule{2-6}
&hxcopd &0.02 &0 &0.03 &0 \\\cmidrule{2-6}
&hxhtn &0.24 &0 &0.25 &0 \\\cmidrule{2-6}
&hxdiabetes &0.09 &0 &0.11 &0 \\\cmidrule{2-6}
&hxhld &0.13 &0 &0.14 &0 \\\cmidrule{2-6}
&hx\_anycancer &0.05 &0 &0.01 &0 \\\cmidrule{1-6}
\multirow{11}{*}{\textbf{Medication History}} &seatedbp1 &127.94 &2 &128.59 &0 \\\cmidrule{2-6}
&seatedbp2 &76.73 &2 &77.32 &0 \\\cmidrule{2-6}
&pres\_mtx &0.73 &0 &0.76 &0 \\\cmidrule{2-6}
&pres\_arava &0.05 &0 &0.03 &0 \\\cmidrule{2-6}
&pres\_azulfidine &0.05 &0 &0.07 &0 \\\cmidrule{2-6}
&pres\_plaquenil &0.16 &0 &0.2 &0 \\\cmidrule{2-6}
&pres\_imuran &0 &0 &0 &0 \\\cmidrule{2-6}
&pres\_minocin &0 &0 &0 &0 \\\cmidrule{2-6}
&pres\_pred &0.33 &0 &0.3 &0 \\\cmidrule{2-6}
&nonpresNSAIDs\_use &0.38 &0 &0.41 &0 \\\cmidrule{2-6}
&NSAIDs\_use &0.22 &0 &0.26 &0 \\\cmidrule{1-6}
\multirow{6}{*}{\textbf{Assessment}} &tender\_jts\_28 &11.9 &0 &10.52 &0 \\\cmidrule{2-6}
&swollen\_jts\_28 &8.25 &0 &7.43 &0 \\\cmidrule{2-6}
&md\_global\_assess &53.03 &2 &51.7 &1 \\\cmidrule{2-6}
&pt\_global\_assess &52.57 &0 &53.86 &0 \\\cmidrule{2-6}
&di &0.6 &0 &0.5 &0 \\\cmidrule{2-6}
&pt\_pain &53.67 &0 &52.74 &0 \\\cmidrule{1-6}
\multirow{4}{*}{\textbf{Lab Test}} &usresultsCRP &13.17 &0 &10.18 &0 \\\cmidrule{2-6}
&usresultsIgA &245.38 &11 &241.54 &6 \\\cmidrule{2-6}
&usresultsIgG &1104.61 &11 &1135.97 &6 \\\cmidrule{2-6}
&usresultsIgM &120.75 &11 &108.41 &6 \\\cmidrule{1-6}
\multirow{2}{*}{\textbf{DAS28}} &DAS28\_CRP\_0M &4.98 &0 &4.82 &0 \\\cmidrule{2-6}
&delta &1.55 &0 &1.4 &0 \\\midrule
\bottomrule
\end{longtable}

\section{Model Interpretation}
\label{model interpretation}
\begin{figure}[!ht]
  \centering
  \includegraphics[width=0.8\linewidth]{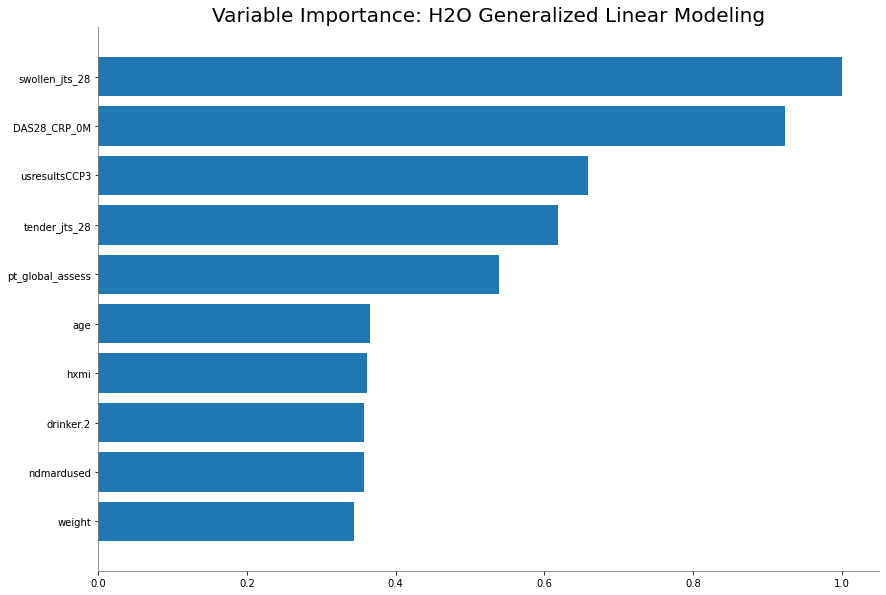}
  \caption{Variable importance plot shows the relative importance of the most important variables in the model (0 to 1).}
  \label{fig:feature importance}
\end{figure}
The variable importance rankings are extracted from the best regression model in the two-stage approach. However, since the stacked ensemble model cannot generate the ranking list, we choose the top predictors of the best base model (GLM) to explain the models. Important features include baseline DAS28, TJC, and CCP3. Besides these features highly related to $\Delta$DAS28, age, weight, drink history, Cyclic Citrullinated Peptide antibody test, and number of DMARDs used are top predictors that contribute to GLM.

\end{appendices}

\end{document}